\documentstyle[preprint,aps]{revtex}
\tightenlines

\begin{document}

\draft
\title{The 3-dimensional Fourier grid Hamiltonian method}
\author{F. Brau\thanks{Chercheur I.I.S.N.} 
and C. Semay\thanks{Chercheur qualifi\'{e} F.N.R.S.}}
\address{Universit\'{e} de Mons-Hainaut, Place du Parc 20,
B-7000 Mons, BELGIQUE}
\date{\today}

\maketitle

\begin{abstract}
A method to compute the bound state eigenvalues and eigenfunctions of 
a Schr\"{o}dinger equation 
or a spinless Salpeter equation with central interaction
is presented. This method is the generalization to the three-dimensional 
case of the Fourier grid Hamiltonian method for one-dimensional 
Schr\"{o}dinger equation. It requires only the evaluation of the potential
at equally spaced grid points and yields the radial part of the 
eigenfunctions at the same grid points. It can be easily extended to the 
case of coupled channel equations and to the case of non-local interactions.
\end{abstract}

\section{Introduction}
\label{sec:intro}

Numerous techniques have been developed to find the eigenvalues and 
eigenvectors of the Schr\"{o}dinger and the spinless Salpeter equations.
In particular, developments of the Hamiltonian in a convenient bases have
been widely used (see for instance Refs.~\cite{fulc94,luch96}). 
The accuracy of the solutions depends
on two parameters: The size of the basis and a characteristic length which
determines the range of the basis states. Upper bounds of the true 
eigenvalues are computed by diagonalizing the corresponding Hamiltonian 
matrix. The quality of the bounds increases with the size of the basis
and, for 
a given number of basis states, there exist a characteristic length which
minimize the value of a particular upper bound. 

\par In the case of a Schr\"{o}dinger equation, other methods
requiring only the evaluation of the potential
at equally spaced grid points, yields directly the amplitude of the 
eigenfunctions at the same grid points \cite{coon81,mars89}. In particular,
the Fourier grid Hamiltonian method \cite{mars89,bali91} appears very 
accurate and simple to handle. This method is variational \cite{ligh85}
and relies on the fact that the kinetic 
energy operator is best represented in momentum space, while the potential 
energy is generally given in coordinate space. 

\par In this paper, we show that this last method can be generalized to
treat the semirelativistic kinetic energy operator, simply by developing
the Fourier grid Hamiltonian method in the 3-dimensional space. 
Consequently, we propose to call our approach, the 3-dimensional 
Fourier grid Hamiltonian method. We focus our attention on the case of 
purely central local potential, but the method can also be applied if the
potential is non-local, or if coupling exist between 
different channels. As explained below, the accuracy of the method depends
on the number of grid points and on the maximal radial distance considered
to integrate the eigenvalue equation. This last parameter is not easy
to calculate without knowing {\em a priori} the wave function, so we propose 
an {\em Ansatz} to determine it. 

\par Our method is outlined
in Sec.~\ref{sec:method}, while Sec.~\ref{sec:dom} presents a convenient
way to compute the domain on which the wave functions are calculated.
Test applications of the method are described in Sec.~\ref{sec:num}, and
a brief summary is given in Sec.~\ref{sec:summary}.

\section{Method}
\label{sec:method}

\subsection{Theory}
\label{sec:theory}

We assume
that the Hamiltonian can be written as the sum of the kinetic
energy $\hat T$ and a potential energy operator $\hat V$. The 
eigenvalue equation for a stationary state is given by
\begin{equation}
\label{tv}
\left[ \hat T + \hat V \right] |\Psi \rangle = E |\Psi \rangle,
\end{equation}
where $\hat T$ depends only on the square of the relative momentum 
$\vec p$ between
the particles, $\hat V$ is a local interaction which depends on
the relative distance,
and $E$ is the eigenenergy of the stationary state. 
This equation is a nonrelativistic 
Schr\"{o}dinger equation if 
\begin{equation}
\label{schro}
\hat T = m_1 + m_2 + \frac{\vec p\,^2}{2 \mu},
\end{equation}
where $m_1$ and $m_2$ are the masses of the particles and
$\mu$ is the reduced mass of the system
(we use the natural units $\hbar = c =1$ throughout the text). 
Equation~(\ref{tv}) 
is a spinless Salpeter equation if 
\begin{equation}
\label{salp}
\hat T = \sqrt{\vec p\,^2 + m_1^2} + \sqrt{\vec p\,^2 + m_2^2}.
\end{equation}
 
\par In configuration space, Eq.~(\ref{tv}) is written
\begin{equation}
\label{tvconf}
\int \left[ \langle \vec r\, |\hat T |\vec r\,' \rangle +
\langle \vec r\, |\hat V |\vec r\,' \rangle \right] 
\langle \vec r\,' |\Psi \rangle\, d\vec r\,' 
= E\, \langle \vec r\, |\Psi \rangle.
\end{equation}
In the following, we only consider the case of a local central potential 
\begin{equation}
\label{vcent}
\langle \vec r\, |\hat V |\vec r\,' \rangle =
V(r)\, \delta(\vec r\, - \vec r\,') \quad
\text{with} \quad r=|\vec r\,|.
\end{equation}
It is then useful to decompose the wave function into its central and
orbital parts
\begin{equation}
\label{focent}
\langle \vec r\, |\Psi \rangle = R_l(r)\, Y_{lm}(\hat r) \quad
\text{with} \quad \hat r = \vec r / r.
\end{equation}
To compute the non-local representation of the kinetic energy operator, we
introduce the basis states $\{ | k \lambda \nu \rangle \}$, 
which are eigenstates of the operator 
$\vec p\,^2$. They are characterized by good orbital quantum numbers 
($\lambda$, $\nu$), obey the relation
\begin{equation}
\label{tp2}
\hat T(\vec p\,^2) | k \lambda \nu \rangle = 
T(k^2) | k \lambda \nu \rangle,
\end{equation}
and satisfy the orthogonality relation
\begin{equation}
\label{ortho}
\langle k' \lambda' \nu'| k \lambda \nu \rangle = 
\delta(k'-k)\, \delta_{\lambda' \lambda}\, \delta_{\nu' \nu}.
\end{equation}
The representation of these states in the configuration space is given by
\begin{equation}
\label{klm}
\langle \vec r\, | k \lambda \nu \rangle = 
\sqrt{\frac{2k^2}{\pi}}\, j_\lambda(kr)\, Y_{\lambda \nu}(\hat r),
\end{equation}
where functions $j_l(kr)$ are spherical Bessel functions. Using the 
completeness relation of basis states 
$\{ | k \lambda \nu \rangle \}$ and Eq.~(\ref{ortho}), we find
\begin{equation}
\label{trrp}
\langle \vec r\, |\hat T |\vec r\,' \rangle =
\int_0^\infty dk\, \frac{2k^2}{\pi}\, T(k^2)\, \sum_{\lambda=0}^\infty
\sum_{\nu=-\lambda}^\lambda j_\lambda(kr)\, j_\lambda(kr')\, 
Y_{\lambda \nu}(\hat r) Y_{\lambda \nu}^\ast (\hat r').
\end{equation}
Introducing the regularized function $u_l(r) = r R_l(r)$, Eq.~(\ref{tvconf})
is written
\begin{equation}
\label{eqrad}
\frac{2}{\pi}\, r \int_0^\infty dr'\, r'\, u_l(r') \int_0^\infty dk\, k^2\,
T(k^2)\, j_l(kr)\, j_l(kr') + V(r)\, u_l(r) = E\, u_l(r).
\end{equation}
This equation is the basis of the 3-dimensional Fourier grid 
Hamiltonian method.

\subsection{Discretization}
\label{sec:discret}

We now replace the continuous variable $r$ by a grid of discrete values
$r_i$ defined by
\begin{equation}
\label{ri}
r_i = i\Delta \quad \text{with} \quad i=0,\, 1,\, \ldots,\, N,
\end{equation}
where $\Delta$ is the uniform spacing between the grid points. 
Regularity at origin imposes $u_l(r_0 = 0) = 0$. For
bound states, we have $\lim_{r \rightarrow \infty} u_l(r) = 0$.
Consequently, we choose to set $u_l(r_N = N\Delta) = 0$. Actually, this last 
condition is not necessary but it does not spoil the accuracy of solutions. 
The normalization condition for the radial wave function is
\begin{equation}
\label{normcont}
\int_0^\infty dr\, \left[ u_l(r)\right]^2 = 1.
\end{equation}
The discretization of this integral on the grid gives
\begin{equation}
\label{norm}
\Delta \sum_{i=1}^{N-1} \left[ u_l(r_i)\right]^2 = 1.
\end{equation}
This corresponds to an integration by trapezoidal rule thanks to the
choice of a vanishing radial wave function at $r=r_0$ and $r=r_N$.

\par The grid 
spacing $\Delta$ in the configuration space determines the grid spacing 
$\Delta k$ in the momentum space. The maximum value of $r$ 
considered being $r_N = N \Delta$, the wave function lives in a sphere
of diameter $2 r_N$	in the configuration space. This length determines
the longest wavelength $\lambda_{\text{max}}$ 
and therefore the smallest frequency 
$\Delta k$ which appears in the $k$-space is
\begin{equation}
\label{dk}
\Delta k = \frac{2 \pi}{\lambda_{\text{max}}} = \frac{\pi}{N \Delta}.
\end{equation}
We have now a grid in configuration space and a corresponding grid in
momentum space
\begin{equation}
\label{gridk}
k_s = s\Delta k = \frac{s \pi}{N \Delta} 
\quad \text{with} \quad s=0,\, 1,\, \ldots,\, N.
\end{equation}
If we note $V_i = V(r_i)$, the discretization procedure
replaces the continuous Eq.~(\ref{eqrad}) by an eigenvalue matrix problem
\begin{equation}
\label{matrix}
\sum_{j=1}^{N-1} H_{ij}\, \phi_j^n = e_n\, \phi_i^n \quad \text{for}
\quad i=1,\, \ldots,\, N-1,
\end{equation}
where
\begin{equation}
\label{hij}
H_{ij} = \frac{2\pi^2}{N^3}\, i\, j \sum_{s=1}^N s^2\, 
T\left( \left( \frac{\pi s}{N \Delta} \right)^2 \right)\,
j_l\left( \frac{\pi}{N}s i \right)\,
j_l\left( \frac{\pi}{N}s j \right)
+ V_i\, \delta_{ij}.
\end{equation}
The ($N-1$) eigenvalues $e_n$ of Eq.~(\ref{matrix}) correspond 
approximately
to the first ($N-1$) eigenvalues of Eq.~(\ref{eqrad}). 
In the case of a potential which 
possesses a continuum spectra, only eigenvalues below the dissociation 
energy are relevant. Other eigenvalues, which form a discrete spectrum of 
positive energies, are spurious and correspond to standing wave 
solutions satisfying $u(r)=0$ at $r=0$ and $r=N\Delta$. The eigenvector
$\phi_i^n$ gives approximately the values of the radial part of the 
$n$th solution of Eq.~(\ref{eqrad}) at the grid points. The eigenvectors
$\phi_i^n$ must be normalized according to Eq.~(\ref{norm}) in order 
that $\phi_i^n \simeq u_l^n(r_i)$. 

\par This method can also be used in the case of a non-local potential. 
If the interaction depends only on the radial variable, then the 
discretization of the action of the potential on the wave function 
gives
\begin{equation}
\label{wrrp}
\int_0^\infty dr'\, W(r,r')\, u(r') \rightarrow 
\Delta \sum_{j=1}^{N-1} W(r_i,r_j)\, u(r_j) \quad \text{for}
\quad i=1,\, \ldots,\, N-1. 
\end{equation}
This corresponds also to an integration by trapezoidal rule thanks to the
choice of a vanishing radial wave function at $r=r_0$ and $r=r_N$.

\par Coupled channels calculations can also be performed with this method.
For instance, let us consider the coupled equations
\begin{equation}
\label{cccont}
\left\{
\begin{array}{ll}
\hat H^{(1)}\, | \phi^{(1)} \rangle + \hat W\, | \phi^{(2)} \rangle  
= &E\, | \phi^{(1)} \rangle \\
\hat W\, | \phi^{(1)} \rangle + \hat H^{(2)}\, | \phi^{(2)} \rangle  
= &E\, | \phi^{(2)} \rangle. 
\end{array}
\right.
\end{equation}
The corresponding discretized equations are 
\begin{equation}
\label{ccdisc}
\left\{
\begin{array}{ll}
\sum_{j=1}^{N-1} \left[ H^{(1)}_{ij}\, \phi^{(1)}_j + W_{ij}\, \phi^{(2)}_j
\right] = &E\, \phi^{(1)}_i \\
\sum_{j=1}^{N-1} \left[ W_{ij}\, \phi^{(2)}_j + H^{(1)}_{ij}\, \phi^{(1)}_j
\right] = &E\, \phi^{(2)}_i. 
\end{array}
\right.
\end{equation}
where $H^{(1,2)}_{ij}$ and $W_{ij}$ are the 3-dimensional Fourier grid 
representation of the interaction operators $\hat H^{(1,2)}$ 
and $\hat W$ respectively. $\phi^{(1,2)}_i$ are approximately 
the values of the radial part of the eigenstates $| \phi^{(1,2)} \rangle$
at grid points $r_i = i\Delta$ for $i=1,\, \ldots,\, N-1$. 

\subsection{Relevance of discretization}
\label{sec:relevance}

As shown in Sec.~\ref{sec:theory},
the 3-dimensional Fourier grid Hamiltonian method relies on the following
relation
\begin{equation}
\label{jj}
\frac{2}{\pi}\, x\,x' \int_0^\infty j_l(kx)\, j_l(kx')\, k^2\, dk =
\delta(x-x').
\end{equation}
The equivalent discrete orthogonality relation on our grid of points is
\begin{equation}
\label{jjdisc}
\frac{2\pi^2}{N^3}\, i\, j \sum_{s=1}^N s^2\, 
j_l\left( \frac{\pi}{N}s i \right)\,
j_l\left( \frac{\pi}{N}s j \right)
= \Delta_{ij}^{(N,l)}.
\end{equation}
One can thus expect that $\Delta_{ij}^{(N,l)} = \delta_{ij}$ for all 
values of $N$ and $l$. Actually, the situation is less favorable.
As it is shown in the appendix, for $l=0$, we have
\begin{equation}
\label{l0}
\Delta_{ij}^{(N,l=0)} = \delta_{ij} \quad \forall N.
\end{equation}
For $l=1$, $\Delta_{ij}^{(N,l=1)} \not= \delta_{ij}$, but
we have verified numerically that 
\begin{equation}
\label{l1}
\lim_{N \rightarrow \infty} \Delta_{ij}^{(N,l=1)} = \delta_{ij}.
\end{equation}
For values of 
$l$ larger than 1, formula (\ref{l1}) is only approximately correct
for small values of $i$ and $j$. Consequently, the accuracy of this method 
becomes poorer when $l$ increases; nevertheless for large enough number 
of grid points, very good results can be obtained.

\section{Domain of integration}
\label{sec:dom}

The accuracy of the eigenvalues and eigenfunctions depends on two 
parameters: The value of $N$ and the value of $r_N$. Obviously, 
for a given value of $r_N$, the accuracy increases with $N$. A 
proper choice for the domain of integration is not evident.
If $r_N$ is too small, incorrect solutions will be found. If this 
parameter is too large, a great number of grid points will be necessary to
obtain stable eigenvalues. In this section, we propose an {\em Ansatz} to 
compute
a suitable value of $r_N$. The idea is to find the radial
distance $r_\epsilon$ for which the radial part $R(r)$ of the eigenfunction 
considered is such that 
\begin{equation}
\label{eps}
\frac{r_\epsilon R(r_\epsilon)}{\max \left[ r R(r) \right]} \leq \epsilon,
\end{equation}
where $\epsilon$ is a number small enough to neglect the contribution of
$R(r)$ for values of $r$ greater than $r_\epsilon$. The eigenfunction 
considered being {\em a priori} unknown, we propose to use a trial wave 
function matching at best the true eigenfunction, at least for
the large $r$ behavior. The value of $r$ 
satisfying the condition (\ref{eps}) for the trial wave function will be 
the value $r_N$ used for the numerical computation.

\par The first step is to find a potential $V_\infty (r)$ which matches 
at best the potential $V(r)$ for $r \rightarrow \infty$. In the following, 
we will consider three different types:
\begin{mathletters}
\label{vinf}
\begin{equation}
\label{vinf:a}
V_\infty (r) = \kappa\ r^p \quad \text{with} \quad \kappa > 0 
\quad \text{and}\quad  p > 0,
\end{equation}
\begin{equation}
\label{vinf:b}
V_\infty (r) = -\frac{\kappa}{r^p} \quad \text{with} \quad \kappa > 0 
\quad \text{and} \quad 0 < p \leq 1,
\end{equation}
\begin{equation}
\label{vinf:c}
V_\infty (r) = -V_0\ \theta(a-r) \quad \text{with} \quad V_0 > 0 
\quad \text{and} \quad a > 0.
\end{equation}
\end{mathletters}
The second step is to choose a trial state $| \lambda \rangle$ which 
depends on one parameter $\lambda$, taken here as the inverse of a distance.
This trial state and the 
eigenstate considered are characterized by similar behaviors for
$r \rightarrow 0$ and $r \rightarrow \infty$. 
The best matching between this state and 
the trial state is obtained by means of the variational
principle. The average value
\begin{equation}
\label{mean}
\langle \lambda | \hat H_\infty | \lambda \rangle = 
\langle \lambda | \hat T + V_\infty(r) | \lambda \rangle
\end{equation}
is then computed and the value of $\lambda$ is determined by the usual
condition
\begin{equation}
\label{lamb}
\frac{\partial \langle \lambda | \hat H_\infty | \lambda \rangle}
{\partial \lambda} = 0.
\end{equation}
In the case of the spinless Salpeter equation, the variational solution is 
computed using the fundamental inequality
\begin{equation}
\label{ineq}
\left\langle \sqrt{\vec p\,^2 + m^2} \right\rangle \leq  
\sqrt{\langle \vec p\,^2 \rangle + m^2 }.
\end{equation}
The radial part $R(r)$ of the trial state
is then analyzed to find the value of $r$ which satisfies the condition
(\ref{eps}).

\par We have remarked that with $\epsilon = 10^{-4}$ it is possible to reach
a relative accuracy better than $10^{-5}$  on eigenvalues, provided $N$ is
large enough ($N \gtrsim 50-100$). A relative accuracy on eigenvalues 
better than
$\epsilon$ can be achieved because the mean value of an observable is
computed using the square of the function $r R(r)$.

\subsection{Ground states}
\label{sec:ground}

We first consider the case of ground states, that is to say states
without vibrational excitation. In the case of
a potential with a large $r$ behavior given by Eq.~(\ref{vinf:a}), we use
harmonic oscillator wave functions as trial states. The radial part is given
by
\begin{equation}
\label{hor}
R(r) =
\sqrt{\frac{2 \lambda^{2 l + 3}}{\Gamma\left(l + \case{3}{2}\right)}}\, 
r^l\, \exp\left(-\case{\lambda^2 r^2}{2}\right). 
\end{equation}
Using procedures (\ref{mean}), (\ref{lamb}) and Eq.~(\ref{ineq}) 
for the spinless Salpeter
equation with potential (\ref{vinf:a}), we find 
\begin{equation}
\label{lamba}
\lambda =
\left[ p\, \kappa \frac{\Gamma\left( l + \case{p+3}{2}\right)}
{\Gamma\left( l + \case{5}{2}\right)} 
\left(
\frac{1}{\sqrt{\left( l + \case{3}{2}\right) \lambda^2 + m_1^2}} +
\frac{1}{\sqrt{\left( l + \case{3}{2}\right) \lambda^2 + m_2^2}}
\right)^{-1}\right]^\frac{1}{p+2}.
\end{equation}
The corresponding relation for the case of a nonrelativistic kinematics is 
obtained with vanishing value for the parameter $\lambda$ in the right-hand 
side of the above formula. The reduced mass of the system appears naturally 
and the equation is no longer a transcendental equation.  

\par If the potential, at great distances, is similar to the potentials
given by Eqs.~(\ref{vinf:b})-(\ref{vinf:c}), the trial states used are the 
bound state Coulomb wave functions. The radial part is written
\begin{equation}
\label{coulr}
R(r) =
\sqrt{\frac{(2 \lambda)^{2 l + 3}}{\Gamma(2 l + 3)}}\, r^l\, 
\exp(-\lambda r).
\end{equation}
The variational calculation for the spinless Salpeter
equation with potential (\ref{vinf:b}), gives 
\begin{equation}
\label{lambb}
\lambda =
\left[ p \, \kappa\, 2^p \frac{\Gamma(2 l + 3 - p)}{\Gamma(2 l + 3)}
\left(
\frac{1}{\sqrt{\lambda^2 + m_1^2}} +
\frac{1}{\sqrt{\lambda^2 + m_2^2}}
\right)^{-1}\right]^\frac{1}{2-p}.
\end{equation}
With the potential (\ref{vinf:c}), we obtain
\begin{equation}
\label{lambc}
\lambda = \frac{1}{2a}
\left[ (2 l +1) \ln (2 \lambda a) -
\ln \left( \frac{\Gamma(2 l +3)}{4 a^2 V_0} 
\left(
\frac{1}{\sqrt{\lambda^2 + m_1^2}} +
\frac{1}{\sqrt{\lambda^2 + m_2^2}}
\right)
\right)
\right].
\end{equation}
Again, the corresponding relations for the case of a nonrelativistic 
kinematics is 
obtained with vanishing value for the parameter $\lambda$ 
under the square roots in the right-hand 
side of the above formulas. The reduced mass of the system appears 
naturally, but Eq.~(\ref{lambc}) remains a transcendental equation.  

Once $\lambda$ is found, it is easy to find $r_N$. Let us 
introduce a dimensionless variable $x_N = 
\lambda r_N$. Using condition (\ref{eps}) with Eqs.~(\ref{hor}) and 
(\ref{coulr}), $x_N$ is given by the transcendental
equation
\begin{equation}
\label{xn}
x_N = \left[ (l+1) \left( \ln \frac{x_N^m}{l+1} + 1 \right) - 
\ln \epsilon^m \right]^\frac{1}{m},
\end{equation}
with $m=2$ in the case of Eq.~(\ref{hor}) and $m=1$ in the case 
of Eq.~(\ref{coulr}). 

\subsection{Vibrational excited states}
\label{sec:vibr}
When the eigenstate considered is characterized by a vibrational excitation
$v$ different from 0, we can use, in principle, the ($v+1$)th 
harmonic oscillator or Coulomb wave function as a trial wave function. But 
such a procedure makes analytical calculation of the optimal $\lambda$ much
more complicated. One knows that the polynomial multiplying the 
exponential term in the ($v+1$)th wave function has degree ($v+l$) in the 
Coulomb case and ($2 v +l$) in the harmonic oscillator case. So we can 
use a trial state with the value of $l$ replaced by an effective orbital
angular momentum $l_{\text{eff}}$ which take into account the highest degree
of the polynomial part of the radial trial state. We have verified that for
potentials with large distance behavior of type (\ref{vinf:a}), it is a 
good approximation to take $l_{\text{eff}} = 2 v +l$. In the case of 
potentials with large distance behavior of types (\ref{vinf:b}) or
(\ref{vinf:c}), it is better to use $l_{\text{eff}} = v +l$.   

\section{Numerical implementation}
\label{sec:num}

We have tested the accuracy of our method with different models found in
the literature \cite{fulc94,luch96,blas90}. In particular, we have
find the same results as those of ref.~\cite{fulc94}, in which a 
Schr\"{o}dinger equation and a spinless Salpeter equation are used with
a potential containing a Coulomb part and a linear part. 
In this section, we only present the results for a Schr\"{o}dinger equation 
with a linear potential and for a spinless Salpeter equation with the
Coulomb potential. 

\par In the model of Ref.~\cite{blas90}, the masses of some meson states are 
simply given by a non-relativistic Hamiltonian with a confinement linear 
potential
\begin{equation}
\label{hblas}
H = m_1 + m_2 + \frac{\vec p\,^2}{2\mu} + ar + C.
\end{equation}
The regularized radial part $u^n(r)$ of the $n$th 
zero orbital angular momentum eigenfunction of this 
Hamiltonian can be written in terms of the Airy function \cite{abra}
\begin{equation}
\label{airy}
u^n(r) = (2\mu a)^{1/6} 
\frac{\text{Ai}\left( (2\mu a)^{1/3}\, r + x_n \right)}
{\sqrt{\int_{x_n}^\infty \text{Ai}^2(x)\, dx }},
\end{equation}
where $x_n$ is the $n$th zero of the airy function. In Fig.~\ref{fig:1},
we show the 8th S-wave 
eigenfunction of Hamiltonian~(\ref{hblas}) for parameters
values: $m_1=m_2=0.300$ GeV, $a=0.1677$ GeV$^2$ and $C=-0.892$ GeV, found
in Ref.~\cite{blas90} (this corresponds to 7th excitation 
of the $\rho$-meson). On 
this figure, the exact function is obtained with formula~(\ref{airy}) and
the numerical one has been computed with a value of $r_N$
calculated with the procedure described in Sec.~\ref{sec:dom} 
for $\epsilon = 10^{-4}$ and with a number
of grid points $N=30$. In these conditions, the 8th eigenvalue
is found with a relative accuracy better than $10^{-4}$. This error 
can be reduced by a factor 10 or more by increasing $N$. The numerical
solution is indistinguishable from the analytical solution to the 
resolution of the figure. If the wave function must be used to compute 
mean values of observables, a greater number of points is obviously 
necessary.

\par None analytical solution of the spinless Salpeter equation with 
Coulomb potential is known. But this equation has been extensively studied
and it is possible to compare the results of our method with results 
from other works. In Table~\ref{tab:1}, we show some eigenvalues of the 
semirelativistic Hamiltonian
\begin{equation}
\label{sscoul}
H = \sqrt{\vec p\,^2 + m_1^2} + \sqrt{\vec p\,^2 + m_2^2} -\frac{\kappa}{r},
\end{equation}
with the parameter values: $m_1 = m_2 = 1$ GeV and $\kappa = 0.456$. The
accuracy of our results are similar of those of Refs.~\cite{fulc94,luch96},
even better for excited states found in Ref.~\cite{luch96}
(the purpose of the work in 
Ref.~\cite{luch96} was not to reach the greatest possible accuracy, but
to demonstrate the feasibility of a method). We have remarked that 
a greater number of grid points is necessary for spinless
Salpeter equation than for Schr\"{o}dinger equation to reach a similar 
accuracy.

\section{Summary}
\label{sec:summary}

The 3-dimensional Fourier grid Hamiltonian method, formulated and tested 
in this paper, appears as a convenient method to find the eigenvalues
and the eigenvectors of a Schr\"{o}dinger or a spinless Salpeter equation.
It has the advantage of simplicity over all the other techniques. In
particular, it requires only the evaluation of the potential at some grid 
points and not the calculation of matrix elements in a given basis. The
method generates directly the values of the radial part of the wave function 
at grid points; they are not given as a linear combination of basis 
functions. Moreover, the extension of the method to the cases of non-local
interaction or coupled channel equations is trivial. 

\par It is worth noting that the method based on the expansion of the wave 
function in basis functions can present some interesting features. In some 
cases, all the matrix elements can be generated from analytic expressions.
Further, the size of the matrices required can be considerably smaller 
(about $20 \times 20$ or $40 \times 40$) \cite{fulc94}.

\par The accuracy of the solutions of the 
3-dimensional Fourier grid Hamiltonian method
can easily be controlled since it
depends only on two parameters: The 
number of grid points and the largest value of the radial distance 
considered
to perform the calculation. A very good estimation of this last parameter 
can be easily determined by using the procedure described above, and the
number of grid points can be automatically increased until a convergence 
is reached for the eigenvalues. The reliability of the method is also 
ensured by its variational character.

\par The method involves the use of matrices of order 
$\left( (N-1) \times (N-1) \right)$, where $N$ is the number of grid
points. Generally, the most time consuming part of the method is the 
diagonalization of the Hamiltonian matrices. This is not a problem for 
modern computers, even for PC stations. Moreover, several powerful 
techniques for finding eigenvalues and/or eigenvectors exist and can be 
used at the best convenience. 
A demonstration program is available via anonymous FTP on 
{\tt umhsp02.umh.ac.be/pub/ftp\_pnt/}.

\acknowledgments
We thank Prof. R. Ceuleneer, Dr F. Michel and Dr Y. Brihaye 
for useful discussions. 
One of us (C.S.) is grateful to Prof. C. Gignoux for providing
useful references.
                                                                                
\appendix
\section*{Orthogonality condition for S-wave state} 

Using the development of spherical Bessel functions in terms of sine and
cosine functions, we have
\begin{equation}
\label{a1}
\Delta_{ij}^{(N,l=0)} = 
\frac{2}{N} \sum_{s=1}^N  
\sin\left( \frac{\pi}{N}s i \right)\,
\sin\left( \frac{\pi}{N}s j \right).
\end{equation}
Replacing the sine function in terms of exponential functions
and using $\sin(0) = \sin(\pi) = 0$, formula 
above becomes
\begin{equation}
\label{a2}
\Delta_{ij}^{(N,l=0)} = 
-\frac{1}{2N} \sum_{s=0}^{N-1}  
\left( e^{i\frac{\pi}{N}si} - e^{-i\frac{\pi}{N}si} \right)\,
\left( e^{i\frac{\pi}{N}sj} - e^{-i\frac{\pi}{N}sj} \right).
\end{equation}
Distributing and using the well-known relation
\begin{equation}
\label{a3}
\sum_{s=0}^{N-1} e^{i\frac{\pi}{N}sj} = \frac{1-e^{i\pi j}}
{1-e^{i\frac{\pi}{N} j}},
\end{equation}
one can obtain Eq.~(\ref{l0}).

\begin{table}
\protect\caption{Energy eigenvalues of the spinless Salpeter equation
with Coulomb potential $V(r) = -\kappa/r$, for the parameter values
$m_1 = m_2 = 1$ GeV and $\kappa = 0.456$. Our results,
for three values of $N$ with a value of $r_N$
calculated with the procedure described in Sec.~\ref{sec:dom} 
for $\epsilon = 10^{-4}$,
are given with the upper bounds 
obtained by the variational methods described in 
Refs.~\protect\cite{fulc94,luch96}.}
\label{tab:1}
\begin{tabular}{cccccc}
State & $N=100$ & $N=200$ & $N=300$ & Ref.~\cite{fulc94} 
& Ref.~\cite{luch96} \\
\hline
1S & 1.9460 & 1.9453 & 1.9451 & 1.9450 & 1.9450 \\
2S & 1.9870 & 1.9867 & 1.9866 & 1.9865 & 1.9868 \\
3S & 1.9944 & 1.9942 & 1.9941 & 1.9941 & 2.0015 \\
4S & 1.9969 & 1.9968 & 1.9967 & 1.9967 & 2.0238 \\
1P & 1.9869 & 1.9869 & 1.9869 & 1.9869 & 1.9875 \\
\end{tabular}
\end{table}

\begin{figure}
\caption{Comparison of exact (solid curve) and numerically computed 
(crosses
surrounded by circles) eigenfunctions for the 7th excitation 
of the $\rho$-meson for the
quark-antiquark Hamiltonian of Ref.~\protect\cite{blas90}. Our computation
is carried out with $N=30$ and an integration
domain determined by the procedure given in Sec.~\protect\ref{sec:dom}
for $\epsilon = 10^{-4}$. 
See the text for further details.}
\label{fig:1}
\end{figure}
 
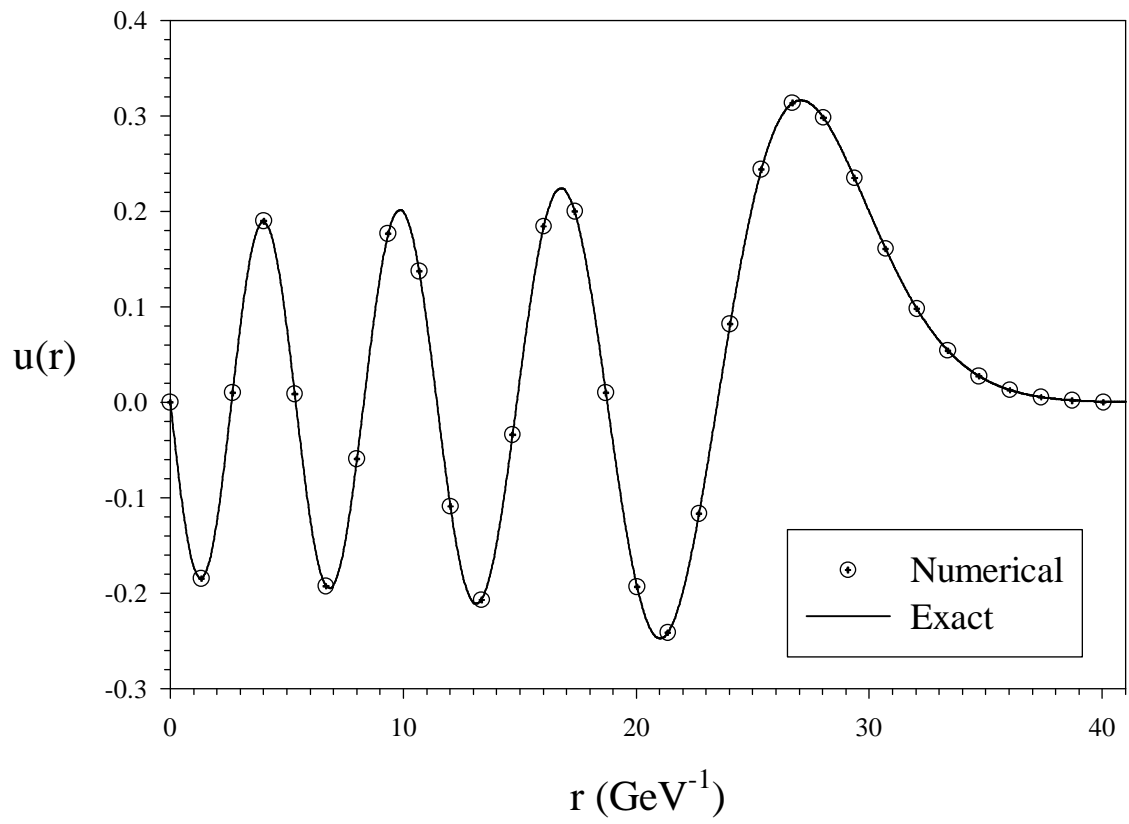
\end{document}